\documentstyle[epsfig]{mn}

\title{Anisotropic inverse Compton emission in the radio galaxy 3C~265}
\author[M. Bondi et al.]
       {M.~Bondi,$^1$ G. Brunetti,$^1$ A. Comastri,$^{2}$
        G. Setti$^{1,3}$ \\
        $^1$Istituto di Radioastronomia, Via Gobetti 101, I-40129, Bologna,
        Italy\\
        $^2$INAF-Osservatorio Astronomico di Bologna, Via Ranzani 1, I-40127.
	Bologna, Italy\\
        $^3$Dipartimento di Astronomia, Universit\`a di Bologna, Via Ranzani 1,
        I-40127 Bologna, Italy\\
}
\begin{document}
\maketitle

\begin{abstract}
We present the results from a {\it Chandra} observation of the powerful radio
galaxy 3C 265. We detect X--ray emission from the nucleus, the radio hotspots 
and lobes. In particular, the lobe X--ray emission is well explained 
as anisotropic inverse Compton scattering of the nuclear photons by the
relativistic electrons in the radio lobes; the comparison between radio
synchrotron and IC emission yields a magnetic field strength a factor 
$\simeq 2$ lower than that calculated under minimum energy conditions.
The X--ray spectrum of the nucleus is consistent with that of a powerful,
strongly absorbed quasar and the X--ray emission of the south-eastern hotspot 
can be successfully reproduced by a combination of synchro-self Compton
and inverse Compton emission assuming a magnetic field slightly lower than 
equipartition.
\end{abstract}


\section{Introduction}
\label{Intro}

The number of powerful radio galaxies (FR IIs) and quasars with detected 
non--thermal X--ray 
emission from the radio lobes and hotspots has steadily increased since the
launch of the {\it Chandra} X--ray observatory. 

Compact X--ray emission from the hotspots is mainly accounted for 
by the synchro--self 
Compton (SSC) mechanism under approximate equipartition conditions, 
although a synchrotron component may be
present in some cases (Hardcastle et al. 2004 and ref. therein). 
On the other hand, 
extended X-ray emission from the lobes
is produced by the inverse Compton (IC) scattering off the 
Cosmic Microwave Background 
(CMB) photons and far-IR/optical photons from the nuclear source.
The IC emission with the CMB is sampling relativistic electrons with
Lorentz factor $\gamma \sim 10^3$, synchrotron emitting at radio wavelengths
in the typical magnetic field of the lobes, while the IC 
emission off the nuclear far-IR/optical photons is powered by 
$\gamma\sim 100-300$ electrons, whose synchrotron emission falls in the 
undetected hundred kHz range. Thus, 
the detection of diffuse X-rays from the IC scattering of the nuclear photons 
provides a unique tool to extend our knowledge of 
the electron spectrum down to lower energies and to constrain the physical 
parameters of the radio lobes \cite{BSC97}.  
Since the IC scattering of
nuclear photons is anisotropic, the far lobe  
of a symmetric double lobed radio galaxy, 
inclined with respect to the plane of the 
sky, should appear more X--ray luminous than 
the near one (Brunetti et al. 1997\nocite{BSC97}; Brunetti
2000\nocite{Brun00}).
Confirmations of this mechanism have come 
from the {\it Chandra} observations of a number
of radio sources (e.g., Brunetti et al.,
2001\nocite{Brun01},2002\nocite{Brun02};
Sambruna et al., 2002\nocite{Samb02}; Carilli 2003\nocite{Cari03}).

3C~265 (z=0.811) is one of the most luminous radio galaxies in the 3CR sample
with two prominent radio lobes extending for about 80 arcsec \cite{Fern93}. 
Optical and infrared observations suggest the presence of a powerful quasar
hidden in the nucleus. By a detailed analysis of the 
spectral energy distribution and optical polarisation, 
di Serego et al. (1996)\nocite{Sere96} have derived an apparent 
de--absorbed magnitude 
$m_V=16.4$ for the central quasar which
corresponds to a bolometric isotropic luminosity in the range 
$L_Q\sim 0.6-2.2\times 10^{47}$ erg s$^{-1}$, depending on the bolometric 
correction assumed \cite{Elvi94}.
This high bolometric luminosity is consistent with the
prominent IR bump emission recently measured by ISOCAM 
($L \geq 2\times 10^{46}$erg s$^{-1}$, Siebenmorgen et al. 2004
\nocite{Sieb04}). The visual extinction toward the nucleus derived from 
these infrared observations is extremely high: $A_V\simeq 64$ and
$N_H\simeq 1.3\times 10^{23}$
cm$^{-2}$ if Galactic conversion values are assumed. 

We observed the radio galaxy 3C~265 with the {\it
Chandra} X--ray observatory to directly test the presence of a
very powerful quasar in the nucleus and of the nuclear photons' IC scattering
scenario outlined above.
Throughout this paper we assume $H_0=70$ km s$^{-1}$ Mpc$^{-1}$ and $q_0=0.5$,
so that 1 arcsec corresponds to 5.9 kpc projected at the source.

\section{Radio And X--ray data}
\subsection{VLA observations}
\label{Radio}

The radio galaxy 3C~265 has been extensively observed with the Very Large
Array (VLA) in the frequency range $1.4-15$ GHz at various angular
resolutions. 
High quality images of 3C~265 have been published by Fernini et al.
(1993)\nocite{Fern93}, whilst our goal is to obtain multifrequency images at 
the same resolution in order to perform a spectral analysis of the source's 
components. For this reason we have reduced and combined 
VLA archive data sets taken with different configurations at four
frequencies (1.4, 4.9, 8.4, and 15 GHz, project codes AF186, AL124, AL200,
AM224 and AV153). 
The data reduction was carried out in the standard way by means of the NRAO
Astronomical Image Processing System (AIPS). 
For the spectral analysis
we obtained two data sets of images: the lower resolution one (circular
beam of 4 arcsec) includes all the 4 frequencies and has been used to
derive the spectral index of the extended lobe emission, the higher resolution 
set (circular beam 1.3 arcsec) has only 1.4, 4.9 and 8.4 GHz images and has 
been used to derive the spectral index of the compact components.

3C~265 has a typical FR~II radio morphology: a weak core, two 
extended lobes (total extension of about 500 kpc)
and bright hotspots at the leading edge of the lobes. The north-western (NW) 
lobe contains two bright regions labelled as C and A by Fernini et al. (1993)
\nocite{Fern93} and both identified as hotspots. 
The south-eastern (SE) hotspot (labelled E) is the brightest component. 

\begin{figure}
\includegraphics[width=4cm,angle=-90]{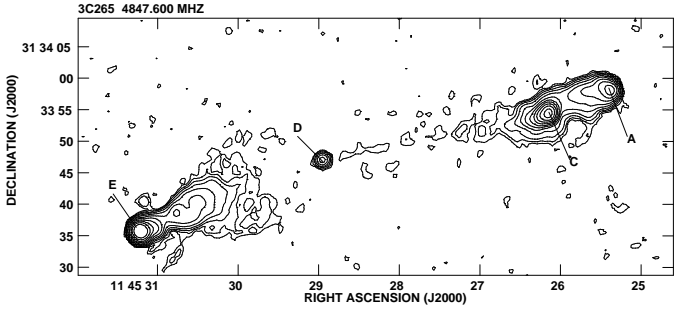}
\caption{\label{3c265_image_radio} Radio image at 4.9 GHz restored with a
circular beam of 1.3 arcsec. Peak flux is 194 mJy/beam. First contour is 0.06
mJy/beam and contour levels are multiple (1, 2, 4, 8 {\ldots})
of the minimum contour level.}
\end{figure}

Figure~\ref{3c265_image_radio} shows the 4.9 GHz image of 3C~265 with 
a resolution
of 1.3 arcsec obtained by combining the A and B configuration data sets.
The off-source rms is 0.022 mJy/beam and together with the components already
identified by Fernini we detect, for the first time, a faint linear feature
west of the nucleus that we identify with the main jet. 

\begin{figure}
\includegraphics[width=4cm, angle=-90]{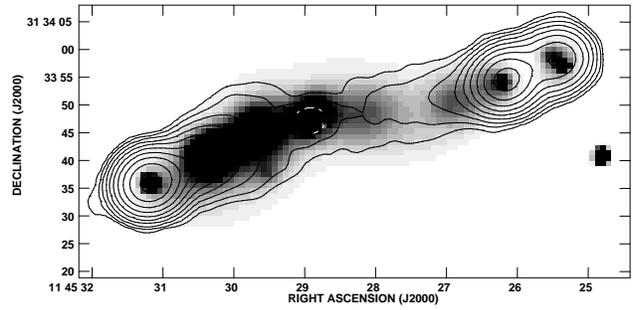}
\caption{\label{3c265_image_x} Contours from the low resolution
(circular beam 4 arcsec) radio image at 4.9 GHz are superimposed on the 
{\it Chandra} image in greys. The X--ray image has been smoothed with a 2.5
arcsecond gaussian kernel.}
\end{figure}

\subsection{Chandra observations}
\label{X-ray}

The target was placed at about $35\arcsec$ from the nominal aimpoint
of the  back illuminated ACIS S3 chip onboard {\it Chandra}
and observed, in Very Faint mode, on 2002 April 24$^{th}$.
The raw level 1 data were re--processed using the latest
version 3.0 of the CXCDS {\tt CIAO} software and filtered 
using a standard grade selection.
The time intervals corresponding to a background level higher than 
the average count rate 
were filtered out leaving about 47.9 ksec of useful data.
The smoothed full band (0.5--7 keV) X--ray image 
is reported in Figure~\ref{3c265_image_x}. 
The radio contours obtained from the low resolution
4.9 GHz radio image are also overplotted.  
The {\it Chandra} position has been found to be offset by 
approximately 0.40 arcsec in
R.A. and 0.13 arcsec in declination, consistently with known {\it
Chandra} astrometric uncertainties. 

A visual inspection of the X--ray image shows several features: 
a relatively bright 
pointlike source coincident with the radio nucleus, enhanced diffuse emission
in the direction of the south-eastern lobe, weaker diffuse emission on the
north-western side of the core, and three X--ray knots spatially
coincident with the radio hotspots.

\section{Results}

In this Section we present the combined results from radio and X--ray 
observations and the modelling of the observed non--thermal emission.
We adopt a simple scenario in which 
the emitting electrons are injected in the 
hotspots' regions and simply age due to synchrotron and IC losses in the 
lobes.
In both hotspots and lobes we thus adopt an electron spectrum which is a 
power law with injection energy index $\delta$ 
which steepens beyond a break energy ($\gamma >\gamma_b$) up to a high 
energy cutoff $\gamma_c$. Throughout the paper, errors are reported at 
the 90\% confidence interval for one interesting parameter.

\subsection{The nuclear source}

The radio core is relatively weak (2.7 mJy at 4.9 GHz) with a spectral index 
($S\propto\nu^{-\alpha}$) $\alpha=0.47\pm 0.08$ between 1.4 and 4.9 GHz 
and $\alpha=0.69\pm 0.16$ between 4.9 and 8.4 GHz.

\begin{figure}
\includegraphics[width=5cm,angle=-90]{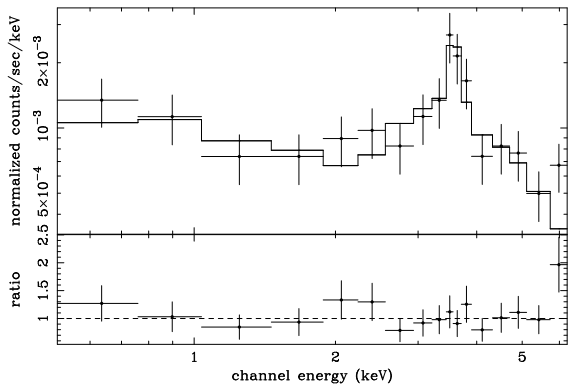}
\caption{\label{nucleus_spectrum}
X--ray spectrum of the nucleus of 3C~265. The spectrum is fitted
with a power-law component+absorption+iron line (see text for details).}
\end{figure}

The nuclear X--ray spectrum has been extracted 
by the cleaned events considering a radius of 1.5 arcsec 
(enclosing about 90\% of the PSF) centered on the maximum of the
X--ray counts distribution. The background spectrum was averaged over several
nearby relatively large regions to minimize small scale fluctuations.
Appropriate response and effective 
area functions were generated using standard tools. 
The spectrum has been grouped with a minimum of 20 counts per bin 
and fitted with the latest version (11.3.0) of {\tt XSPEC}. 

The broad band (0.5--6 keV) spectrum is relatively complex and best 
fitted with an absorbed power law
plus an unabsorbed power law component at low energies 
(Figure~\ref{nucleus_spectrum}).
The 1 keV normalization of the soft component is
about 5\% of the primary one. 

The absorption is intrinsic to the source
and relatively high (N$_{\rm H}=(3.1_{-0.9}^{+1.5})\times 10^{23}$), 
consistent with the visual extinction derived by Siebenmorgen et al. (2004)
\nocite{Sieb04}.
The power law component has a photon index $\Gamma=1.76\pm 0.28$.
The unabsorbed 0.5--7 keV X--ray flux is $\sim$ 2.5 $\times 10^{-13}$ erg 
cm$^{-2}$ s$^{-1}$. The intrinsic X--ray luminosity over the 
broad 0.1--10 keV band of the hidden nucleus is about
5 $\times 10^{44}$ erg s$^{-1}$ (with a maximum value of about a few $10^{45}$
erg s$^{-1}$ at the 90\% confidence level for $\Gamma$).
A narrow iron line at 6.4 keV is clearly detected in the X-ray spectrum of
the core, even if
the counting statistic at high energy is not such to allow a detailed 
analysis of the line properties. Fixing the rest--frame energy at 6.4 keV 
and the line width at the instrumental resolution the rest frame 
equivalent width is 490$\pm$340 eV.

The observed high column density, the iron line and the intrinsic core
luminosity are a direct and unambiguous signature of the presence of a powerful
hidden quasar obscured by a dense, dusty torus in the nucleus of 3C~265.

\subsection{The lobes}

The low resolution (4 arcsec) radio images clearly trace the extended emission
of the lobes from the hotspots back to the nucleus. 
The low frequency (178--750 MHz) spectral index of the source is 
$\alpha=0.95\pm 0.1$. 
The SE lobe shows a steepening of the spectral index:
$\alpha=1.59\pm 0.06$ (1.4 -- 4.9 GHz) and 
$\alpha=1.93\pm 0.25$ (8.4 -- 14.9 GHz).
The flux density ratio between the SE and NW lobes derived at 1.4 GHz is 
$\sim 2.7$.

Extended emission in the X--ray image 
is clearly detected on both side of the nucleus 
of 3C~265. The X--ray emission is cospatial with the radio lobes 
thus strongly pointing to a non-thermal origin.
The X--ray distribution is clearly asymmetric, with the X--ray SE lobe 
brighter than the NW one. This asymmetry, combined
with the presence of a faint radio jet in the NW lobe,
suggests that most of the extended X--rays may be due to the
IC scattering of the nuclear photons.
Source spectra have been extracted
using appropriate response and effective area functions.
The instrument response has been also corrected for the
degradation in the {\tt ACIS} quantum efficiency using the latest version
of the {\tt ACISABS} tool.
Integrating over the whole source area and subtracting the contribution 
from the core and hotspots, we derive $200\pm 15$ net counts 
(0.5-2 keV band) for the total extended emission of which 140 
are due to the SE lobe.
The 0.5--2 keV fluxes are $1.2\times 10^{-14}$~erg~cm$^{-2}$~s$^{-1}$
and $0.5\times 10^{-14}$~erg~cm$^{-2}$~s$^{-1}$ for the 
SE and NW lobes, respectively, so that
the flux ratio is $\sim 2.4$, roughly the same
as the brightness ratio. 
The total extended emission can be fitted with a power law with 
$\Gamma=2.35\pm 0.6$ (plus Galactic absorption) which
is consistent, within the errors, with the radio synchrotron spectrum. 
The 0.5-2 keV flux of $1.7\times 10^{-14}$~erg~cm$^{-2}$~s$^{-1}$ 
corresponds to a rest-frame luminosity of $3.3\times 10^{43}$~erg~s$^{-1}$.
The IC scattering off the
nuclear photons would roughly outweigh that of the CMB photons at
angular distances (in arcsecond) from the nucleus :

\begin{equation}
R_{\rm arc} < 4\times L^{1/2}_{46}(1-\mu)$$
\label{raic}
\end{equation}

where $L_{46}$ is the (isotropic) nuclear luminosity in units of $10^{46}$
erg s$^{-1}$ in the far-IR to optical band and 
$\mu=\cos(90\pm \theta_{ax})$
with $\theta_{ax}$ the angle between the radio axis and the plane of the sky
(negative for the near lobe).
With $L_{46}$ being about half of the bolometric luminosity 
$L_Q\sim 0.6-2.2\times 10^{47}$ erg s$^{-1}$ derived
by di Serego et al. (1996)\nocite{Sere96},
and $\theta_{ax} \simeq 10^\circ - 20^\circ$,
it is found that the major contributor to the X--ray extended emission of the
far lobe is the anisotropic IC with the far-IR/optical photons from the 
central quasar (Eq.\ref{raic}), as indeed suggested by the asymmetric
X--ray brightness distribution.
The faintness of the radio jet and the absence of depolarization 
asymmetry in the lobes \cite{Fern01} are strong indicators that 
the radio source axis must lie almost on the plane of the sky justifying the
range of $\theta_{ax}$ adopted.

\begin{figure}
\includegraphics[height=6.0cm]{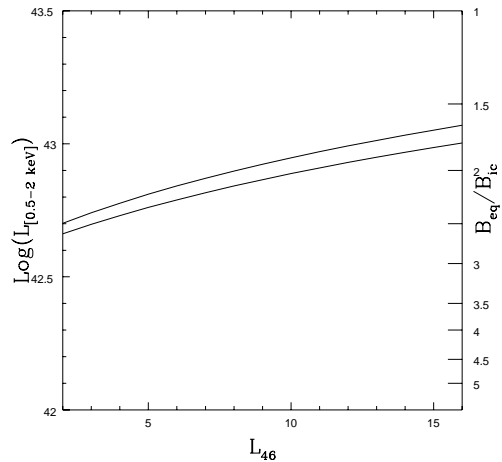}
\caption{\label{lum_trg1p2} 
The two curves show the logarithm of the X--ray predicted IC luminosity of 
the lobes (left vertical scale; the observed lobes' luminosity
on this scale is 43.5) and the ratio between the magnetic field 
derived from equipartition and that derived from the observed X--ray IC 
emission (right vertical scale) versus the QSO nuclear luminosity in the
far-IR to optical band, in units of 
$10^{46}$ erg s$^{-1}$, for two different half opening angles of the
radiation cone ($\theta_C=35^\circ$ and $\theta_C=45^\circ$ for the lower 
and upper curve respectively), 
assuming an orientation of the radio axis $\theta_{ax}\simeq 17^\circ$ 
with respect to the plane of the sky.}
\end{figure}

We calculate the X--ray emission from the radio lobes
by combining the contribution from the anisotropic 
IC scattering of the nuclear photons and that of the isotropic
scattering of the CMB photons.
The anisotropic scattering is computed 
by integrating the Brunetti (2000) equations 
with $\delta=2.85$ and $\gamma_b > 10^3$, so that the results are insensitive
to $\gamma_c$, $\gamma_b$ and $\Delta\delta$.
We also
assume that the nuclear photons are 
isotropically emitted within a cone, of half opening angle 
$\theta_C=35^\circ$ or $\theta_C=45^\circ$, with the axis 
coincident with the radio axis.
Finally, in our calculations the spatial distribution of the
emitting electrons is modelled 
with two half-ellipsoid volumes, whose semiaxis are $31\times 8$
arcsec and $40\times 7$ arcsec for the SE and NW lobe respectively.
The volume averaged equipartition magnetic field in the lobes, 
as derived from the 178 MHz radio flux by making use of
minimum energy formulae with a low energy cut--off
$\gamma_{\rm min}=50$ (e.g., Brunetti et al. 1997\nocite{BSC97}),
is about 46 $\mu$G.

\begin{figure}
\includegraphics[height=6.0cm]{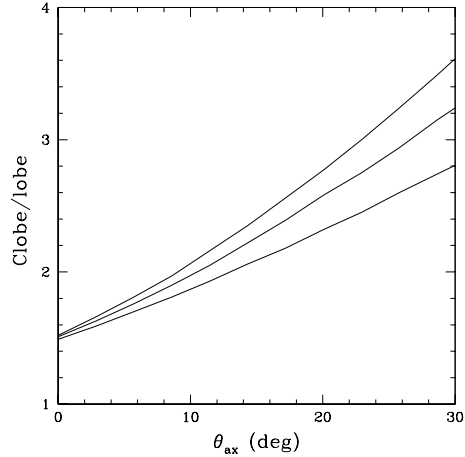}
\caption{\label{rapporto_q}  Counter-lobe/near-lobe X--ray flux ratio
versus $\theta_{ax}$,
the angle between the source axis and the plane of the sky. Different curves
correspond to different quasar luminosities (L$_{46}=7-10-13$ erg
s$^{-1}$ from bottom to top); $\theta_C=45^\circ$ has been adopted.
}
\end{figure}

As can be seen in Figure~\ref{lum_trg1p2} we find that, under equipartition 
conditions ($\gamma_{\rm min} =50$), the predicted IC luminosity is about a 
factor of 4 lower than the observed one (for $L_{46}=10$). 
To account for this difference the average field intensity ($B_{ic}$) 
should be a factor $\simeq 2$ lower than the 
equipartition value ($B_{eq}$).
A moderate departure from equipartition conditions
is consistent with a number of recent {\it Chandra} and XMM findings
(e.g. Brunetti et al. 2002\nocite{Brun02}; Hardcastle et al.
2002\nocite{Hard02};
Comastri et al. 2003\nocite{Coma03}; Grandi et al. 2003\nocite{Gran03};
Isobe et al. 2002\nocite{Isob02}).
Given the above assumptions, we derive the IC X--ray flux ratio of the
two lobes as a function of $\theta_{ax}$ for different 
quasar luminosities and cone apertures (Figure~\ref{rapporto_q}).
The predicted flux asymmetry is found to be in the range $2-3$, for 
$\theta_{ax}\simeq 15^\circ - 20^\circ$, in very
good agreement with the observations, provided that the 
far lobe is the SE one (we stress that this ratio is independent
on the assumed magnetic field intensity in the lobes).
In principle the observed X-ray asymmetry can
also be reproduced via the IC scattering of CMB photons
by assuming an {\it ad hoc} asymmetry in the energetics
of the two radio lobes, such that
$B_{eq}/B_{ic} \sim 3.2$ and $\sim 2.3$
(corresponding to an energy ratio between
electrons and field of $\sim 100$ and $\sim 30$)
are required for the SE and NW lobes,
respectively.
The fact that this {\it ad hoc} model requires an
energetics $\sim 2.5$ times larger than the anisotropic
IC model, the detection of the faint radio jet in the
NW lobe (indicating that the SE lobe is further away
from us in agreement with the geometry required
by the anisotropic IC model) and the morphology of the
X-ray emission (stronger in the innermost regions of the
SE lobe and not strictly correlated with the radio
brightness) allow us to consider the anisotropic IC
model a more realistic interpretation.

\subsection{The south-eastern hotspot}

At the highest resolution (0.35 arcsec, Fernini et al. 1993\nocite{Fern93}) 
the SE hotspot is clearly elongated transversely to the radio axis, a 
possible indication of a bow shock--like 
structure or of the presence of multiple
sub-components. Using the 1.3 arcsecond resolution maps
we have performed a radio spectral analysis of the hotspot region coincident
with the X--ray emission
deriving $\alpha=1.07\pm 0.06$ between 1.4 and 4.9 GHz and  
$\alpha=1.17\pm 0.11$ between 4.9 and 8.4 GHz.
A reasonable fit to the radio hotspot consists of two gaussian components: 
a compact region (FWHM $1.0\times 0.87$ arcsec) and a more extended 
component (FWHM $1.9\times 1.1$ arcsec) responsible for the elongated shape 
of the hotspot in the higher resolution radio images.

The SE hotspot is clearly detected in the {\it Chandra} image with $16\pm 4$ 
net counts. 
The peak of the X--ray emission is within 0.1 arcsec from the radio peak,
although no distinction can be made between the two radio
components; the 0.5-2 keV flux is 
$\sim 1.2\times 10^{-15}$ erg cm$^{-2}$ s$^{-1}$.
X--ray emission is also detected from both hotspots A \& C in the NW lobe but, 
due to the limited statistics, here we focus on the SE hotspot only.

Due to the structure of the hotspot we consider 3 contributions to
the predicted X--ray emission:
\begin{itemize}
\item
SSC emission from the most compact region (FWHM $1.0\times 0.87$ arcsec)
in the hotspot;
\item
IC emission produced by scattering of the electrons of the compact region
with the synchrotron photons from the more extended hotspot region (FWHM
$1.9\times 1.1$ arcsec);
\item
IC emission with the CMB photons from both the compact
and extended hotspot components;
\end{itemize}

\begin{figure}
\includegraphics[height=6.0cm]{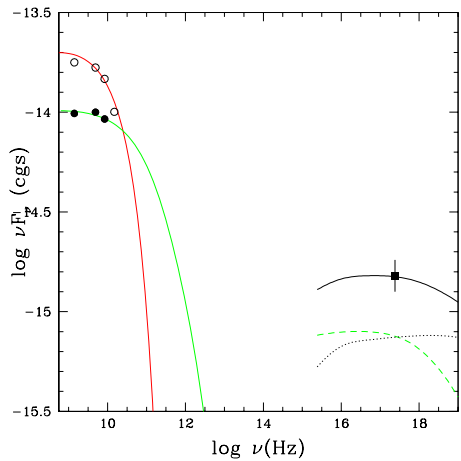}
\caption{\label{hs-s} The radio and X--ray data of the hotspot E in 3C
265 compared with the predicted X--ray emission. The open circles
are the total hotspot flux with a representative synchrotron
model (not used in the IC calculations) and the filled ones the flux from the
compact component only with the proper synchrotron spectrum. The square gives
the observed X-ray flux.
The dashed line represents the
SSC emission from the compact component and the additional IC emission with the 
synchrotron photons from the more extended hotspot region. The dotted line is 
the IC emission with the CMB photons from
both the hotspot's components. The continuous line is the sum of the dashed and
dotted lines. The X--ray spectra have been calculated assuming a
magnetic field a  factor of 1.7 smaller than the equipartition values.}
\end{figure}

In modelling the inverse Compton emission we follow Brunetti et al. (2002)
\nocite{Brun02}, assuming $\delta=2.95$, $\nu_c=6.7\times 10^{12}$ Hz,
$\nu_b=1.3\times 10^{11}$ Hz and $\gamma_{low}<\gamma_* <<\gamma_b$.
We note that the results are insensitive to the choice of $\gamma_{low}$ and
$\gamma_*$.
The comparison between the model predictions and data is reported in
Figure~\ref{hs-s}. The radio data show both the total hotspot spectrum,
derived from the low resolution images, and the compact component.
The IC spectra have been calculated assuming a magnetic field of 70 $\mu$G
and 90 $\mu$G in the extended and compact component respectively. These
values are a factor 1.7 lower than the equipartition magnetic fields derived
in these regions ($B\simeq 150$ $\mu$G in the compact region and 
$B\simeq 120$ $\mu$G in the more extended component). 
Although more complex modelling is adopted here,
the results are consistent with an independent estimate of the SSC
emission from the hotspot recently made by Hardcastle et al. 
(2004)\nocite{Hard04}.

\section{Summary}

We have presented the results from a relatively deep {\it Chandra} 
observation of the powerful radio galaxy 3C 265, supported by complementary
radio data.
Evidence of a powerful quasar hidden in the nucleus of this radio
galaxy has been previously gathered by optical spectro--polarimetry and 
infrared observations .

The combination of a very powerful nucleus and of a prominent
radio structure with strong synchrotron emission allow us to 
test the mechanism of IC scattering with the nuclear photons.

The main results of our analysis can be summarized as follows:

\begin{enumerate}
\item
A powerful, strongly absorbed (N$_{\rm H}=(3.1_{-0.9}^{+1.5})\times 10^{23}$)
X--ray nucleus is detected; the de-absorbed power law photon spectrum 
($\Gamma=1.76\pm 0.28$) and luminosity (about
5 $\times 10^{44}$ erg s$^{-1}$) are consistent with those of
a powerful quasar as conjectured by previous indirect measurements
in other bands.
 
\item
Extended X--ray emission is clearly detected in the 
{\it Chandra} image with total rest frame luminosity of
$3.3\times 10^{43}$~erg~s$^{-1}$.
This emission 
is nicely cospatial with the radio lobes and its power law 
spectrum ($\Gamma=2.35\pm 0.6$)
consistent with that of
the radio synchrotron emission, strongly indicating
a non--thermal (IC) origin. The lobes emission is asymmetric,
the NW lobe being relatively fainter.

\item
The extended emission is mostly contributed by the anisotropic 
IC scattering of the far-IR/optical photons from the powerful central
quasar. A clear signature of this process comes from the
detected asymmetric X--ray emission of the lobes, implying that
the the radio source axis is inclined by an angle $\theta_{ax}$ 
with respect to the plane of
the sky, the fainter NW lobe being nearer to us. This is confirmed
by our detection of a one-sided radio jet pointing in the direction
of the NW lobe.

\item
By a detailed modelling of the anisotropic IC scattering
we are  able to reproduce all the observed properties of the 
extended emission with $\theta_{ax}\simeq 17^\circ$ and a half opening angle of
the quasar radiation cone $\theta_C=35^\circ$--$45^\circ$,
under the assumption that the average magnetic field
intensity is a factor $\simeq 2$ lower than the equipartition value. 

\item
The SE hotspot, showing a somewhat complex radio structure, is clearly 
detected by {\it Chandra}. Its X--ray emission can be accounted for 
by a contribution of the compact component, via the SSC mechanism and 
the IC scattering off
the synchrotron photons from the more extended 
radio component, plus a contribution from the IC scattering off the CMB 
photons of both
components, under the assumption of an average magnetic field strength
a factor 1.7 lower than the equipartition value.
\end{enumerate}

\section*{Acknowledgments}
We thank the referee, Martin Hardacastle, whose helpful comments have 
improved the paper.
This research has made use of the NASA/IPAC Extragalactic
Database (NED) which is operated by the Jet Propulsion
Laboratory, California Institute of Technology, under contract with the
National Aeronautics and Space Administration.
The authors acknowledge partial support by the ASI contract I/R//057/02 and
the MIUR grants Cofin 03-02-7534 and 03-02-23.

\end{document}